\documentclass{eas}
\usepackage{graphicx}
\begin{document}

\title{Dynamics of substructures in warm dark-matter cosmologies}
\runningtitle{Arnold \etal : Substructures in warm dark-matter cosmologies}
\author{Bastian Arnold}\address{Institut f\"ur Astronomie, Universit\"at Wien, T\"urkenschanzstra\ss e 17, A-1180, Wien, Austria}
\author{Alexander Knebe}\address{Astrophysikalisches Institut Potsdam, An der Sternwarte 16, D-14482, Potsdam, Germany}
\author{Chris Power}\address{Centre for Astrophysics and Supercomputing, Swinburne University of Technology, Hawthorn 3122, Victoria, Australia}
\author{Brad K. Gibson}\address{Centre for Astrophysics, University of Central Lancashire, Preston, PR1 2HE, UK}
\begin{abstract}
We performed cosmological simulations based upon both a cold dark matter (CDM) and a warm dark matter (WDM) model. The focus of our investigations lies with selected spatial and kinematic properties of substructure halos (\emph{subhalos}) orbiting within host halos, that form in both dark-matter cosmologies. We aim at using the dynamics of the subhalos as a probe of the respective cosmology.
\end{abstract}
\maketitle
\section{Introduction}
The currently favoured concept of structure formation in our evolving Universe, the $\Lambda$CDM model, is successfully applied on the clustering of matter at intermediate to large scales. However, cosmological $N$-body simulations are lacking good agreement with observations at galactic scales and below. The most promising small-scale modification to this model can be provided when one allows the dark matter to be warm rather than being cold. This assumes that dark matter was relativistic at the time of decoupling and hence entails a non-negligible \emph{free-streaming scale}, $\lambda_f$, corresponding to a \emph{filtering mass}, $M_f$ (\cite{Colin2008}). Thus, the power spectrum of structure formation in the CDM model must be corrected for scales below $\lambda_f$ (\cite{Bardeen1986}) to allow for suppression of structures.
\section{The simulations}
Our simulation suite was carried out using the publicly available code \texttt{MLAPM} (\cite{Knebe2001}), which integrates Poisson's equation on an adaptive grid. We focus on the formation and evolution of galaxy-cluster sized dark-matter halos containing about one million particles. The mass resolution and force resolution are given by $1.6\times 10^8h^{-1}$M$_\odot$ and $\sim 2h^{-1}$kpc, respectively. At redshift $z=45$, an initially homogeneous distribution of $512^3$ particles is placed in a simulation box having a volume of ($64h^{-1}$Mpc)$^3$. After collapsing the respective closest $8$ particles into more massive ones, this lower resolution of $128^3$ particles is evolved until $z=0$. The particles within $5R_{vir}$ of identified dark-matter halos are then restored to their former mass resolution. For the surrounding layers of particles a conservative criterion is chosen, such that the jump in refinement between two adjacent cells is not greater than one (\cf\ \cite{Tormen1997}).
\newline
The only difference between the two dark-matter models is given by their respective power spectrum. With an adopted warmon mass $m_{WDM}=0.5$keV, a matter density $\Omega_0=0.3$, and a Hubble constant $h=0.7$, we obtain a free-streaming scale $\lambda_f=2.46h^{-1}$Mpc and a corresponding filtering mass $M_f=6.4\times 10^{11}h^{-1}$M$_\odot$ (for details see Knebe \etal\ \cite{Knebe2008}). All structures with masses $M<M_f$ are suppressed in the WDM model.
\section{The results}
For the sake of brevity, the results being presented here are only a selection from Knebe \etal\ (\cite{Knebe2008}) and can be read in more detail there.
\subsection{Distinct subhalo populations}
Simulations, which are based upon CDM models, suggest a significant fraction of subhalos on the outskirts of present-day host halos (\emph{e.g.} \cite{Warnick2008}). This ``backsplash'' population once resided inside the virial radius of its respective host halo, but is found outside at $z=0$. We discovered such a backsplash subset in our WDM model, too (\cf\ fig.~\ref{fig:1}). Nevertheless, the reduced abundance of backsplashed subhalos reflects their lower central densities with respect to their CDM counterparts, which in turn implies a more likely disruption within a host halo.
\subsection{Mass loss}
It is common to both dark-matter models that the mass loss the subhalos suffer over the entire life time of the underlying host halo is the greater the deeper they penetrate into its potential well. But irrespective of their halo-centric distance, the mass loss of subhalos is enhanced in the WDM model compared to CDM (\cf\ fig.~\ref{fig:2}), which again confirms the assumption that WDM-generated subhalos are less concentrated than subhalos in the CDM model (\cite{Avila-Reese2001}; \cite{Knebe2002}).
\subsection{Spatial anisotropy}
We detected the spatial distribution of subhalos by measuring the angles between their position vectors and the eigenvector corresponding to the major axis of their host halo. It turned out, that the spatial anisotropy of backsplashed subhalos is more pronounced than for bound subhalos supporting the fact that backsplash halos tend to be on more radial orbits. Infalling subhalos in the WDM cosmology are on average more confined to the major axis of the respective host halo (\cf\ fig.~\ref{fig:3}) and so are very low-mass subhalos ($M_{halo}<10^{-4}M_{host}$).
\subsection{Relative-velocity distribution}
\begin{figure}[!ht]
\noindent
\begin{minipage}[ht]{.49\linewidth}
\flushleft
\includegraphics[width=\linewidth]{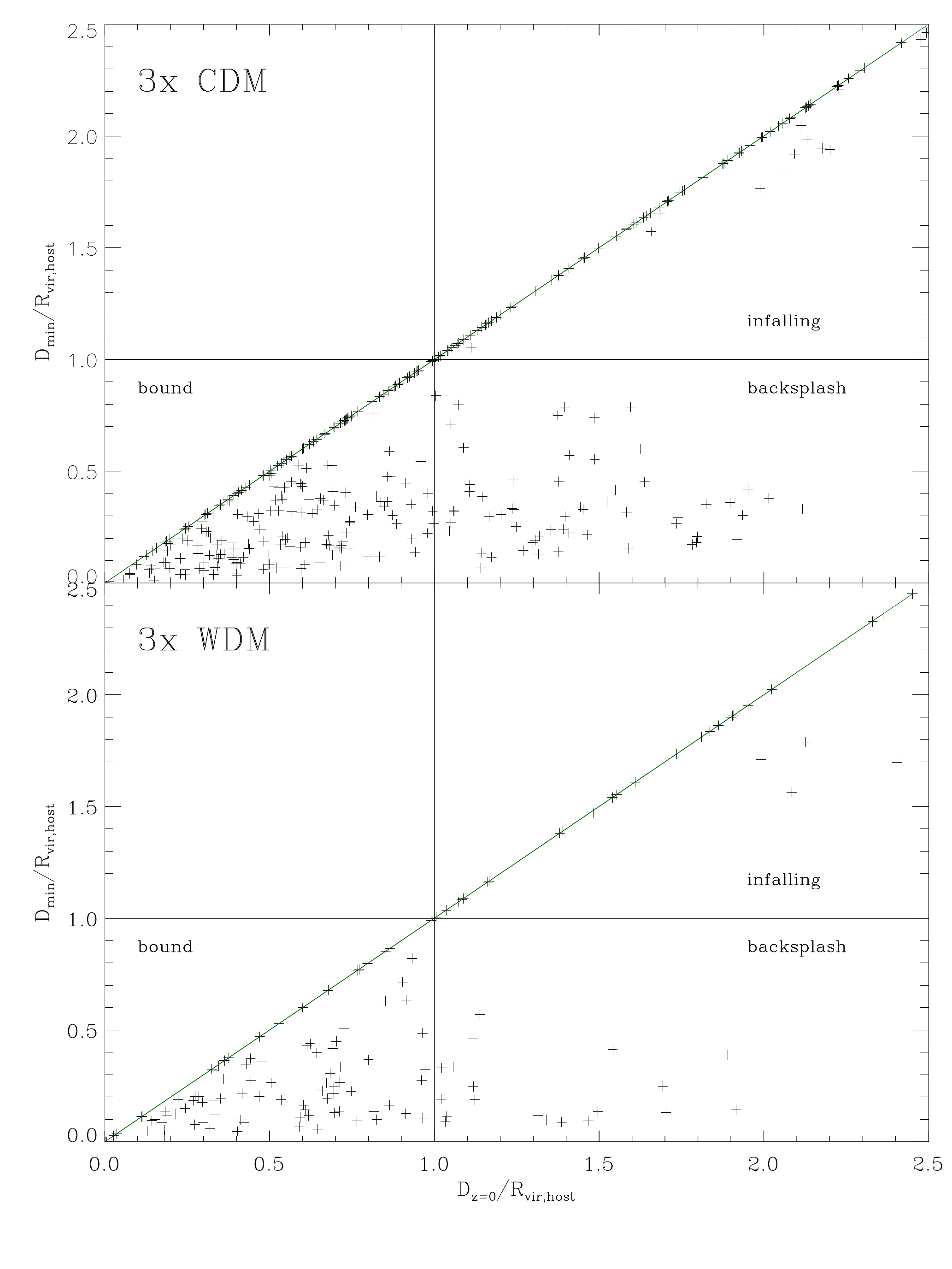}
\caption{Plotting the minimum distance with respect to the present-day halo-centric distance divides the subhalos into evolutionary distinct populations.}
\label{fig:1}
\end{minipage}
\begin{minipage}[ht]{.49\linewidth}
\flushright
\includegraphics[width=\linewidth]{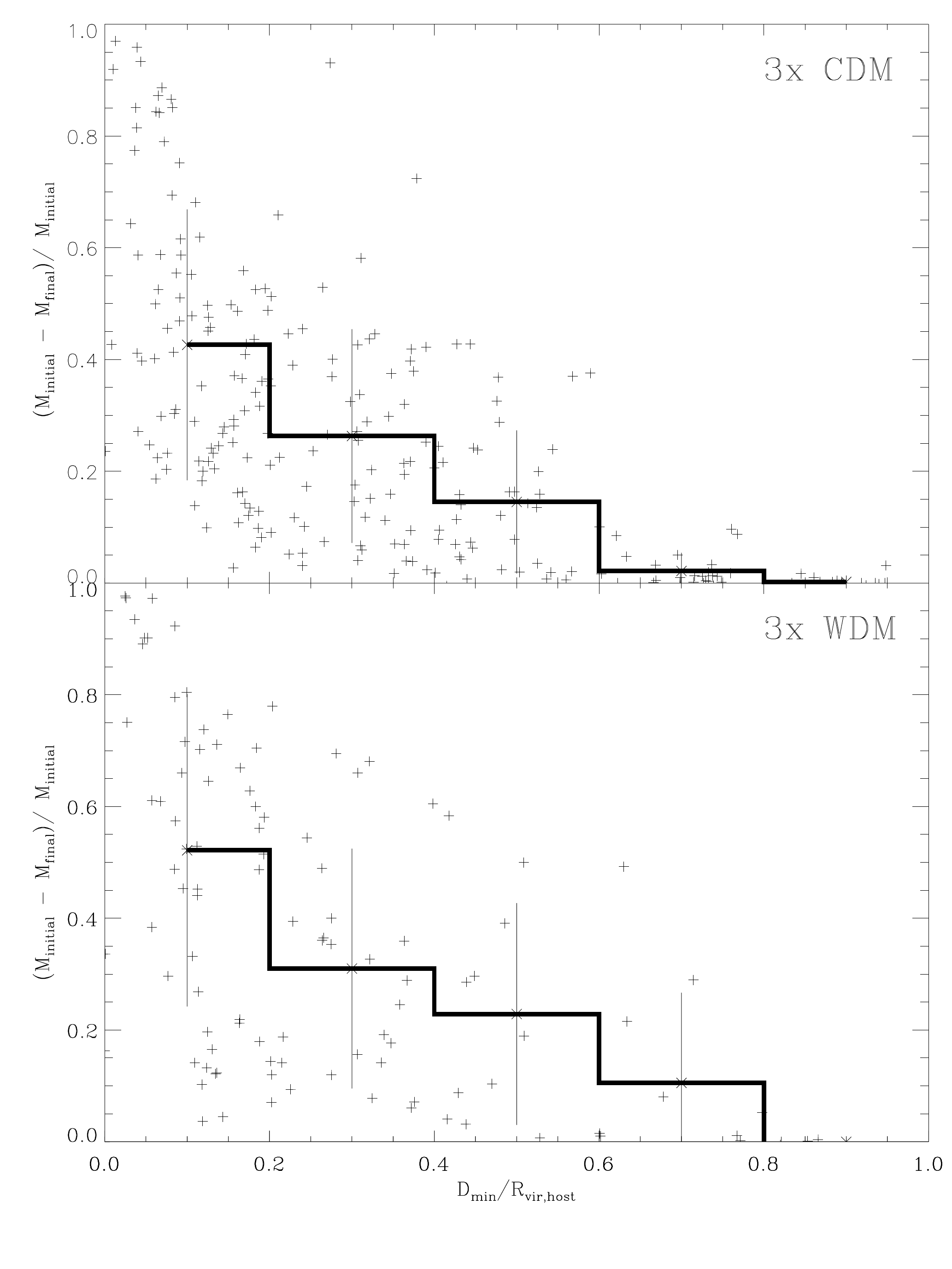}
\caption{The mass loss of bound subhalos against their minimum halo-centric distance is measured over the life time of the respective host halo.}
\label{fig:2}
\end{minipage}
\begin{minipage}[ht]{.49\linewidth}
\includegraphics[width=\linewidth]{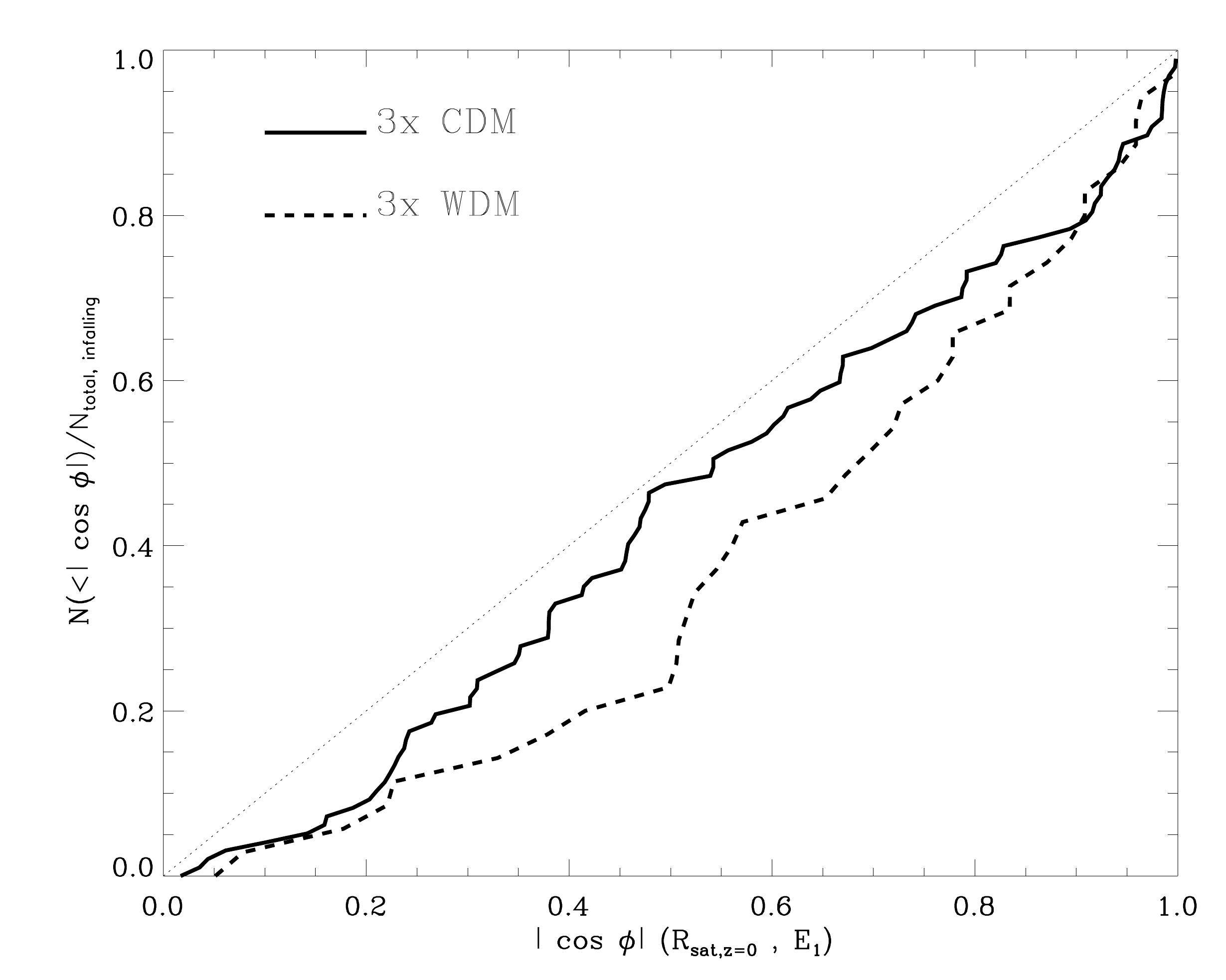}
\caption{The cumulative distribution that scans the level of isotropy of all infalling subhalos. The dotted line denotes an isotropic distribution.}
\label{fig:3}
\end{minipage}
\begin{minipage}[ht]{.49\linewidth}
\includegraphics[width=\linewidth]{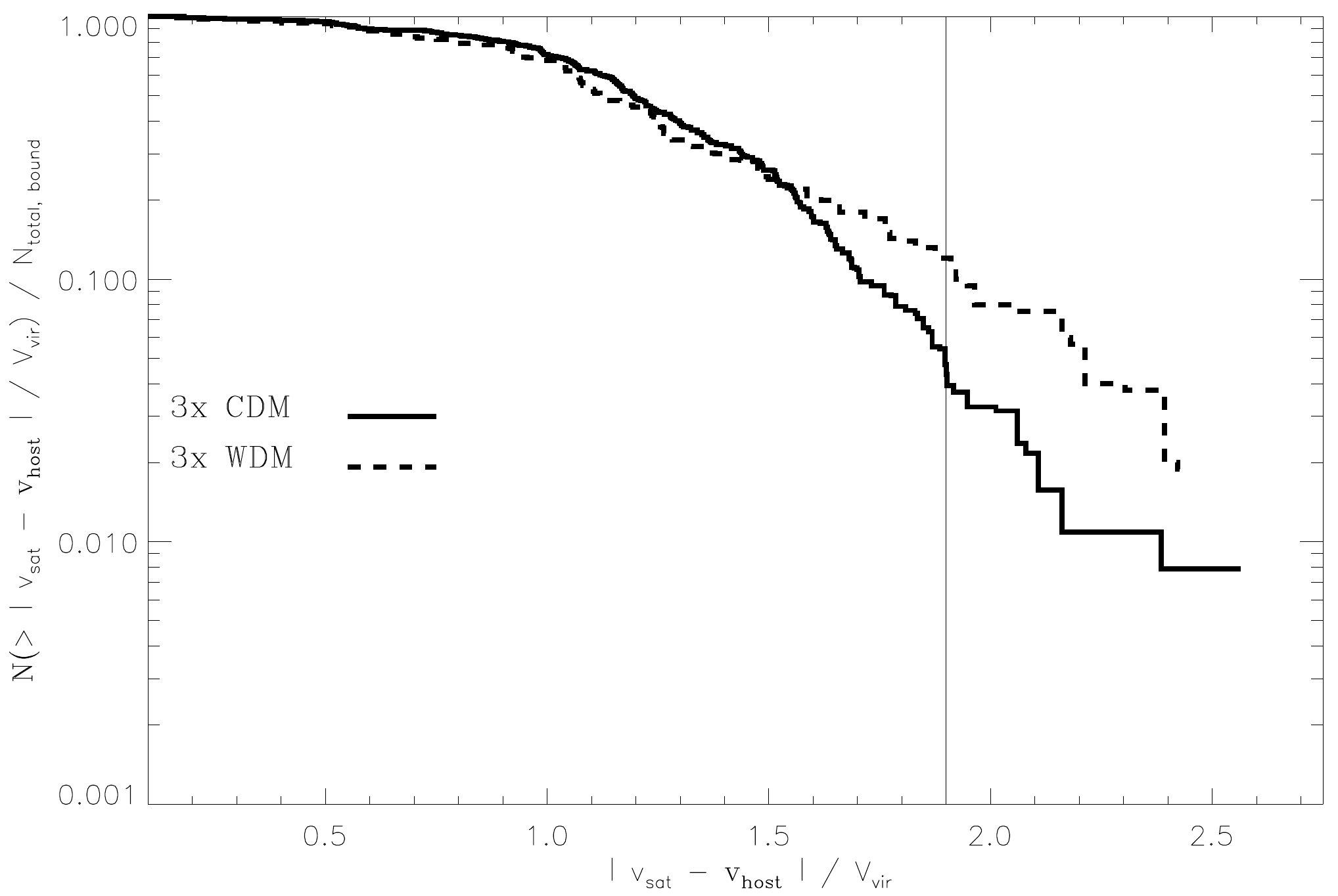}
\caption{The number fraction of bound subhalos with a higher relative velocity than a given one. The velocities are normalized to the circular velocity of the respective host halo at its virial radius.}
\label{fig:4}
\end{minipage}
\end{figure}

The bound subhalo population in the WDM model is slightly faster than its counterset in the CDM cosmology (\cf\ fig.~\ref{fig:4}). Thus, high-speed encounters like the observed ``Bullet Cluster'' (\cite{Markevitch2004}; \cite{Milosavljevic2007}) are more likely explained within the WDM model. This extremely high-velocity merger between two galaxy clusters is denoted by the vertical line in fig.~\ref{fig:4} at a normalized collision speed $|v_{sat}-v_{host}|/V_{vir}\sim 1.9$. However, the fastest subhalo was detected in the CDM simulations giving a clue on stronger effect from dynamical friction in the WDM cosmology.
\section{Discussion and conclusions}
We have investigated and compared the properties of substructures orbiting in galaxy-cluster sized host halos in both a WDM and a CDM cosmology. Both models are able to generate a ``backsplash'' population and a similar anisotropic distribution of subhalos with respect to the major axis of the underlying host halo. On the other hand, we confirmed the enhanced mass loss via tidal stripping for low-mass subhalos in the WDM model compared to their counterparts in the CDM model, as was shown before (\emph{e.g.} \cite{Colin2008}). Infalling subhalos in the WDM model have on average higher velocities than the corresponding subset in the CDM model.
\newline
We conclude that the spatial and kinematic properties of subhalos do not allow us to clearly differentiate between both dark-matter cosmologies at $z=0$.
\end{document}